# Harnessing oversampling in correlation-coded OTDR

Ruolin Liao,[1] Ming Tang,[1,*] Can Zhao,[1] Hao Wu,[1] Songnian Fu,[1] Deming Liu,[1] and Perry Ping Shum[2]

[1]*Wuhan National Laboratory for Optoelectronics (WNLO) and National Engineering Laboratory for Next Generation Internet Access System, School of Optics and Electronic Information, Huazhong University of Science and Technology, Wuhan 430074, China*
[2]*School of Electrical and Electronics Engineering, Nanyang Technological University, 637553, Singapore*

**Abstract:** Pulse coding is an effective method to overcome the trade-off between signal-to-noise ratio (SNR) and spatial resolution in optical-fiber sensing systems based on optical time-domain reflectometry (OTDR). However, the coding gain has not been yet fully exploited. We provide a comprehensive theoretical analysis and experimental validation of the sampling criteria for correlation-coded OTDR, showing that the coding gain can be further improved by harnessing the oversampling. Moreover, the bandwidth-limited feature of the photodetector can also be utilized to select the sampling rate so that additional SNR enhancement is obtained. We believe this principle could be applied to any practical OTDR-based optical-fiber sensing technology, and serve to update existing systems based on correlation-coded OTDR in a straightforward manner at a relatively low cost.

## 1. Introduction

Optical time-domain reflectometry (OTDR) is an important distributed optical-fiber sensing technique, in which backscattered light is acquired to obtain the characteristics along the fiber under test. However, since the Rayleigh scattering coefficient is very small, the backscattered light is very weak as compared to the probing light, leading to a poor signal-to-noise ratio (SNR) in OTDR systems. Although increasing the pulse width can improve the SNR, it also deteriorates the spatial resolution. One solution to this trade-off is to apply pulse coding. In essence, pulse coding increases the injected energy to improve the SNR, while the spatial resolution, which is determined by the bit width, remains the same.

In direct detection OTDR systems, coding schemes can be classified into two groups: linear combination codes and correlation codes [1]. Within the first group, the Simplex coding scheme is the most representative and has been demonstrated effective in various OTDR-based fiber sensing systems [2–4]. The biorthogonal code is another linear combination code that has proved to be useful [5]. However, all coding schemes in this group suffer from the same problem, namely that enhancing the coding gain implies extending the code length, which results into a remarkable increase in measurement time. This is especially troublesome in long-range or dynamic/instantaneous measurements. On the other hand, within the second group of coding schemes, complementary-correlation codes are particularly suitable for OTDR systems due to their perfect autocorrelation property. The most representative among them is the Golay code [6]. The number of sequences in correlation codes is independent of the code length; hence, the number of periods needed to perform a complete measurement remains constant. Other correlation coding schemes have also been developed to further reduce measurement time [7] or enhance coding gain [8].

Interestingly enough, as illustrated in the following sections, the coding gain for a correlation coding scheme can be further enhanced by simply choosing an appropriate analog-to-digital converter (ADC) sampling rate after photodetection. This method does not require any change in the transmitter, which means it can be easily applied to existing OTDR systems with a correlation coding scheme, thus enhancing their performance at a relatively low cost. However, this technique is not compatible with linear combination codes, as will be proved below.

In this paper, we present a comprehensive analysis of correlation-coded OTDR, taking into consideration the ADC sampling rate and photodetector (PD) bandwidth. We will prove that the coding gain can be further improved by setting a sampling rate that is higher than the bit rate of the input light pulses, i.e., by oversampling. Moreover, the coding gain is also affected by the PD bandwidth, and extra SNR enhancement can be obtained by setting the sampling rate within a specific range relevant to the noise characteristics of the PD. The results of the experiments performed confirm these conclusions.

## 2. Theory

### A. The correlation relationship in OTDR

The light source used in direct detection OTDR is incoherent, so an OTDR system can be regarded as a discrete linear time-invariant system. The OTDR response $f_n$ is defined as the backscatter response to an optical pulse signal, usually a rectangular pulse. When certain coding is applied to OTDR, a number of identical pulses arranged according to the unipolar binary code sequence $c_n$ are injected into the fiber, each of which results in the very same backscatter response with different time delay. The total response $y_n$ is their superposition. When the acquisition card records data with a sampling interval equal to the pulse width, $y_n$ is given by [6]

$$y_n = c_n * f_n, \qquad (1)$$

where $*$ represents the convolution operation.

Correlation is one of the spread-spectrum techniques that can provide measurements with improved SNR without sacrificing response resolution. It is implemented by correlating the detected signal with the probe signal. Note that both convolution and correlation are distributive and associative (however, convolution is commutative whereas correlation is not). Thus, one obtains

$$c_n \otimes y_n = c_n \otimes (c_n * f_n) = (c_n \otimes c_n) * f_n, \quad (2)$$

where $\otimes$ denotes correlation. If the autocorrelation of the probe signal is a delta function, $f_n$ can be accurately recovered as

$$c_n \otimes y_n = \delta_n * f_n = f_n. \quad (3)$$

In this case, rather than the probe signal itself (which may be long and energetic), it is the autocorrelation of the probe signal that determines the response resolution. However, binary sequences with zero sidelobes are not known to exist. One approach to pursue the synthesis of zero sidelobes is to sum the autocorrelations of a group of complementary codes. For example, two $L$-element sequences form a Golay complementary pair, if the sum of their autocorrelations is zero for all nonzero shifts [9]:

$$A_n \otimes A_n + B_n \otimes B_n = 2L\delta_n. \quad (4)$$

By probing the system under test with Golay codes $A_n$ and $B_n$, the output signals are expressed as

$$\begin{cases} y_{An} = A_n * f_n \\ y_{Bn} = B_n * f_n \end{cases}. \quad (5)$$

Correlating these outputs with their respective probe sequences and summing the results, one obtains

$$z_n = A_n \otimes y_{An} + B_n \otimes y_{Bn} = (A_n \otimes A_n + B_n \otimes B_n) * f_n = 2Lf_n. \quad (6)$$

Thus, we can precisely recover $f_n$ from $y_{An}$ and $y_{Bn}$.

The code words consist of bipolar elements, which cannot be implemented in a direct detection scheme, so it is necessary to transmit the unipolar version of these codes as the probe sequences. The construction of unipolar codes is achieved through

$$\begin{cases} u_n^A = (1+A_n)/2 \\ \bar{u}_n^A = (1-A_n)/2 \\ u_n^B = (1+B_n)/2 \\ \bar{u}_n^B = (1-B_n)/2 \end{cases}. \quad (7)$$

Transmitting these four unipolar sequences to the system, and then subtracting the received signals between each pair of unipolar codes, we get

$$\begin{cases} u_n^A * f_n - \bar{u}_n^A * f_n = A_n * f_n = y_{An} \\ u_n^B * f_n - \bar{u}_n^B * f_n = B_n * f_n = y_{Bn} \end{cases}. \quad (8)$$

With (6) and (8), the fiber response is recovered by using unipolar codes.

The effect of pulse coding in OTDR systems is evaluated through the coding gain, which is defined as the SNR improvement with respect to conventional averaged OTDR, when the same measurement time is used. When white noise is considered the main source of noise in the OTDR system, it can be found from [6] that the coding gain is $\sqrt{L}/2$, where $L$ is the length of the Golay code.

### B. The effect of oversampling

The discussion above was based on the hypothesis that the ADC sampling interval is the same as the pulse width. However, this is not an indispensable condition, and the sampling interval can be much shorter than the pulse width, as long as the pulse width is an integer multiple of the sampling interval—otherwise the correlation process cannot be correctly carried out. Such multiple relation may affect the decoding process; hence, it is necessary to investigate the impact of oversampling in correlation-coded OTDR systems.

Let us assume that the pulse width is $m$ times the sampling interval, where $m$ will be referred to as the oversampling ratio. This is equivalent to replicate $m$ times each element in the code words. If we denote the pulse width by $T$, then the time interval between adjacent sampling points is $T/m$; however, the pulse width is not changed. Taking, for example, a bipolar Golay pair of length 2 and $m = 2$, the replication is as follows:

$$\begin{cases} A_n = [1 \ 1 \ 1 \ -1] \\ B_n = [1 \ 1 \ -1 \ 1] \end{cases} \rightarrow \begin{bmatrix} 1 & 1 & 1 & 1 & 1 & 1 & -1 & -1 \\ 1 & 1 & 1 & 1 & -1 & -1 & 1 & 1 \end{bmatrix}. \quad (9)$$

Then the result of (4) is no longer a delta function, but a triangle function. More specifically, the sum of the autocorrelations of a pair of $m$-times oversampled Golay codes with original code length $L$ is a triangle function with a peak value of $2Lm$ and an FWHM (full width at half maximum) of $m$:

$$A_n \otimes A_n + B_n \otimes B_n = \sum_{k=-m}^{m} 2L(m-|k|)\delta(n-k) = 2Lmq_n^{(m)}, \quad (10)$$

where

$$q_n^{(m)} = \sum_{k=-m}^{m}\left(1-\frac{|k|}{m}\right)\delta(n-k) = \begin{cases} 1-|n|/m, & -m \leq n \leq m \\ 0, & \text{otherwise} \end{cases} \quad (11)$$

is the unit triangle function, with a peak value of 1 and an FWHM of $m$. Obviously, this function does not depend on the code length. Figure 1 depicts the correlation result when $m$ equals 2 and the code length is 2. The sampling interval is $T/m$, so the FWHM of $q_n^{(m)}$ in the continuous domain is the same as the width of a single bit in the code sequence, which also means they have the same energy.

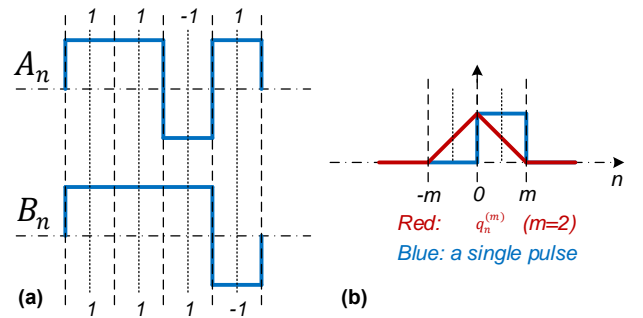

Fig. 1. (a) Code word of a Golay pair when the code length is 2 and $m = 2$. (b) Comparison between a single rectangular pulse (blue) and the equivalent triangular pulse (red).

Equation (6) becomes

$$z_n = 2Lmq_n^{(m)} * f_n' = 2Lmf_{n(tri)}. \quad (12)$$

Note that $f_n'$ in (12) is the OTDR response to a single rectangular pulse of width $T/m$, (rather than $T$, as for $f_n$). Then $f_{n(tri)}$ can be seen as the OTDR response to the unit triangular pulse defined in (11). From (12), we see that oversampling is equivalent to probing the fiber with a triangular pulse rather than the conventional rectangular one. These two probing pulses have the same FWHM and energy, which implies that the system spatial resolution remains the same and the corresponding OTDR responses, i.e., $f_n$ and $f_{n(tri)}$, should have roughly the same intensity.

Now let us turn to the coding gain analysis. The signal voltage at the PD is proportional to the light intensity, so it is reasonable to use signal voltage and noise voltage to calculate the SNR. Only white noise is taken into consideration, so the noise voltage has a normal distribution with mean value 0 and standard deviation $\sigma$. Hence, a noise signal $N^{(k)}$ is added to each of the four received raw signals, so Eq. (8) is revised as

$$\begin{cases} \left(u_n^A * f_n + N_n^{(1)}\right) - \left(\bar{u}_n^A * f_n + N_n^{(2)}\right) = y_{An} + N_n^{(1)} - N_n^{(2)} \\ \left(u_n^B * f_n + N_n^{(3)}\right) - \left(\bar{u}_n^B * f_n + N_n^{(4)}\right) = y_{Bn} + N_n^{(3)} - N_n^{(4)} \end{cases}, \quad (13)$$

whereas, according to (6), Eq. (12) becomes

$$z_n = 2Lm f_{n(tri)} + A_n \otimes \left(N_n^{(1)} - N_n^{(2)}\right) + B_n \otimes \left(N_n^{(3)} - N_n^{(4)}\right). \quad (14)$$

The last two terms in (14) represent noise, and they will be grouped together and denoted by $N^{(total)}$ in the following discussion. Note that $A_n$ and $B_n$ here have $Lm$ elements due to oversampling. Since noise is evaluated by its root-mean-square (RMS) value, our objective is to find out the RMS of $N^{(total)}$:

$$N_n^{(total)} = A_n \otimes \left(N_n^{(1)} - N_n^{(2)}\right) + B_n \otimes \left(N_n^{(3)} - N_n^{(4)}\right)$$
$$= \sum_{k=1}^{Lm} \left\{ A_k \left(N_{k+n}^{(1)} - N_{k+n}^{(2)}\right) + B_k \left(N_{k+n}^{(3)} - N_{k+n}^{(4)}\right) \right\}. \quad (15)$$

$A_n$ and $B_n$ are made up by $-1$'s and $1$'s, so the RMS of $N^{(total)}$ is given by

$$\sigma_n^{(total)} = \sqrt{D(N_n^{(total)})} = \sqrt{Lm \times \left\{ \left(\sigma^2 + \sigma^2\right) + \left(\sigma^2 + \sigma^2\right) \right\}}$$
$$= 2\sqrt{Lm}\sigma, \quad (16)$$

where the operator $D$ denotes variance. Thus, the SNR of the Golay code correlation process is given by

$$SNR_c = \frac{2Lm f_{n(tri)}}{\sigma_n^{(total)}} = \sqrt{Lm} \frac{f_{n(tri)}}{\sigma}. \quad (17)$$

In order to obtain the coding gain, it is necessary to compare the SNRs of the Golay-coded and single-pulse probe signals under similar measurement conditions, i.e., for the same peak power and measurement time. Thus, we should perform four single-pulse measurements and calculate the final result as the average. This process is expected to enhance the SNR by a factor 2, so we have

$$SNR_p = 2 \frac{f_n}{\sigma}. \quad (18)$$

As discussed before, $f_n$ and $f_{n(tri)}$ should have roughly the same intensity, so the coding gain is given by

$$\text{coding gain} = \frac{SNR_c}{SNR_p} = \frac{\sqrt{Lm}}{2}. \quad (19)$$

From (19), it is clear that oversampling contributes to a further enhancement of the coding gain that has not been reported before. Oversampling actually turns a code sequence of $L$ bits into one of $Lm$ bits and, from this point of view, the coding gain conforms to the expression for correlation-coded OTDR without oversampling. This phenomenon occurs not only for the Golay code, but also for any other complementary-correlation coding schemes, such as CCPONS [8]. Moreover, in the case of coded-OTDR based on linear combination coding schemes such as the Simplex code [10], as there is no correlation in the decoding process, oversampling does not affect the coding gain.

### C. Noise analysis for a bandwidth-limited PD

Equation (16) is based on the assumption that the noise signals are uncorrelated, i.e. $E\{N_m^{(a)} N_n^{(b)}\} = 0, (a,b=1,2,3,4; m \neq n)$. In discrete sampling, this is true when the PD has an infinite bandwidth, or when the sampling rate is lower than the PD bandwidth. In OTDR systems, the bandwidth of the PD is often chosen to be higher than the bit rate of input light pulses, but when oversampling is applied, the sampling rate may exceed the PD bandwidth. Hence, the noise should be regarded as bandwidth-limited white noise, which has the following features:

$$E\left\{N_n^{(a)}\right\} = 0; E\left\{N_n^{(a)} N_{n+k}^{(b)}\right\} = 0; E\left\{N_n^{(a)} N_{n+k}^{(a)}\right\} = R_N(k);$$
$$E\left\{\left(N_n^{(a)}\right)^2\right\} = R_N(0) = \sigma^2; \ (a,b=1,2,3,4; a \neq b), \quad (20)$$

where $R_N(k)$ is the autocorrelation function of noise. From Eq. (20), noise samples are uncorrelated for different $a$ and $b$, so the four components in $N^{(total)}$ can be separately calculated and summed to obtain the final result. Taking e.g. the first component, since the mean value of noise is always 0, we have

$$D\left\{\sum_{k=1}^{Lm} A_k N_{k+n}^{(1)}\right\} = E\left\{\left(\sum_{k=1}^{Lm} A_k N_{k+n}^{(1)}\right)^2\right\}$$
$$= \sum_{i=1}^{Lm} \sum_{j=1}^{Lm} A_i A_j E\left\{N_{i+n}^{(1)} N_{j+n}^{(1)}\right\} = \sum_{i=1}^{Lm} \sum_{j=1}^{Lm} A_i A_j R_N(j-i). \quad (21)$$

The range of $(j - i)$ in Eq. (21) is from $-(Lm - 1)$ to $(Lm - 1)$; then, we can simplify this equation by replacing $j$ with $(i + k)$:

$$D\left\{\sum_{k=1}^{Lm} A_k N_{k+n}^{(1)}\right\} = \sum_{k=-(Lm-1)}^{Lm-1} \sum_{i=1}^{Lm} A_i A_{i+k} R_N(k)$$
$$= \sum_{k=-(Lm-1)}^{Lm-1} R_A(k) R_N(k), \quad (22)$$

where $R_A(k)$ represents the autocorrelation function of $A_n$. Similarly, the amended version of Eq. (16) is given by

$$\sigma_n^{(total)} = \sqrt{2 \sum_{k=-(Lm-1)}^{Lm-1} \left(R_A(k) + R_B(k)\right) R_N(k)}$$
$$= \sqrt{4Lm\left(\sigma^2 + 2\sum_{k=1}^{m-1} q_k^{(m)} R_N(k)\right)}. \quad (23)$$

In Eq. (23), the quantity $R_A(k) + R_B(k)$ is the same as in Eqs. (4) and (10). For ideal white noise, the autocorrelation function $R_N(k)$ is a $\delta$ function, so the second term in Eq. (23) is 0 and Eq. (23) reduces to Eq. (16). Similarly, when there is no oversampling, Eq. (23) also coincides with Eq. (16) according to Eq. (4), which means that the bandwidth-limited feature of the PD will not exert any extra effect in nonoversampling correlation coding schemes. However, when the sampling rate is higher than the bandwidth of the PD—the situation we are mostly interested in—it is not evident whether the second term (which in the following discussion will be called *extra term*, for convenience) in Eq. (23) will be positive or negative. If it is negative, then Eq. (23) implies that an extra SNR enhancement can be obtained. Next, we will prove that it is possible to achieve this by choosing a proper sampling rate.

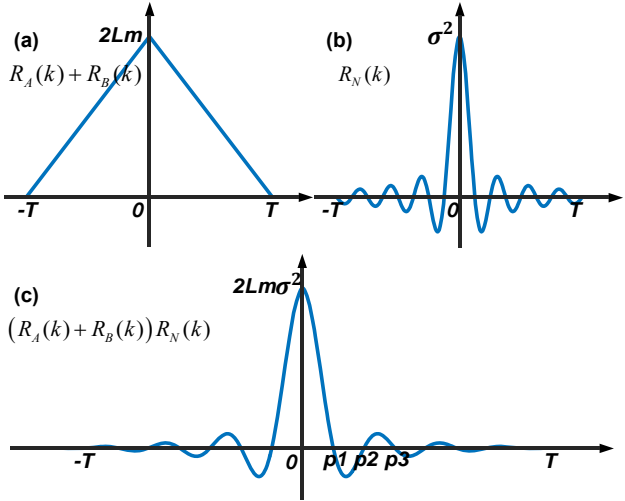

Fig. 2. Estimation of $\sigma_n^{(total)}$. (a) Sum of the autocorrelations of $A_n$ and $B_n$. (b) Autocorrelation of noise. (c) Their product. The points $p1$, $p2$, and $p3$ are the first three zero-crossings.

The spectrum of noise is determined by the frequency response of the PD, which is usually a lowpass filter. For convenience, a rectangular window is adopted as an ideal lowpass filter to represent the noise spectrum, so the autocorrelation of noise is a sinc function, as shown in Fig. 2(b). This assumption does not affect our conclusions. On the other hand, we have already demonstrated that the sum of the autocorrelations of the Golay codes is a triangle function; see Fig. 2(a). The product of these autocorrelations is displayed in Fig. 2(c), where the first three zero-crossing points are indicated. For an ideal lowpass filter, $p1$, $p2$, and $p3$ are $1/2B$, $1/B$, and $3/2B$, respectively, where the bandwidth $B$ is the equivalent noise bandwidth of the PD, usually larger than the frequency response bandwidth. For a real PD, $p1$ is also $1/2B$, but the other two points may not follow the same multiple relationship. This curve is defined in the continuous time domain, so the result of Eq. (23) is obtained by sampling it and summing up all the samples.

Our primary concern is whether the extra term in Eq. (23) is positive or negative. As we know, the zero-crossing points of the sinc function have good periodicity. Further, the oscillation of this function attenuates quickly, an attenuation further accelerated by the triangular window. When the sampling interval is set at a valley of the curve in Fig. 2(c), the extra term will be negative, resulting in a higher coding gain than the theoretical value $\sqrt{Lm}/2$. This extra SNR enhancement will be maximum if the sampling interval is set between $p1$ and $p2$, especially at $(p1+p2)/2$. However, for the same reason, if the sampling interval is set between $p2$ and $p3$, the bandwidth-limited character of the PD will reduce the SNR enhancement. Since the curve attenuates rapidly, continuing to increase the sampling interval will make the extra term approach zero, corresponding to the case in which the sampling rate is within the bandwidth of the PD, so the noise can be seen as ideal white noise. On the other hand, when the sampling interval is smaller than $p1$, the extra SNR enhancement will be smaller. When the sampling interval is sufficiently small, i.e., the oversampling ratio $m$ is sufficiently large, the sum in Eq. (23) will approach an integral. Note that an oversampling ratio $m$ means that the time interval between adjacent sampling points is $T/m$, so we have

$$\sum_{k=-(Lm-1)}^{Lm-1} \left(R_A(k)+R_B(k)\right)R_N(k) = 2Lm \sum_{k=-(m-1)}^{m-1} q_k^{(m)} R_N(k)$$
$$= \frac{2Lm^2}{T}\left(\sum_{k=-(m-1)}^{m-1} q_k^{(m)} R_N(k)\cdot\frac{T}{m}\right) = \frac{2Lm^2 C}{T}. \quad (24)$$

Again, the sum in Eq. (24) approaches an integral when $m$ is sufficiently large. The value of such integral depends only on the autocorrelation function of the PD noise and thus can be regarded as a constant, denoted by $C$. Then the coding gain is given by

$$\text{coding gain} = \frac{2Lmf_{n(tri)}/\sqrt{4Lm^2C/T}}{2f_n/\sigma} = \frac{\sigma}{2}\sqrt{\frac{LT}{C}}. \quad (25)$$

Equation (25) implies the coding gain will eventually reach an upper limit when increasing the sampling rate.

In conclusion, the sampling rate should be set between $1/p1$ and $1/p2$ in order to get extra SNR enhancement by utilizing the bandwidth-limit feature of the PD. For the real PDs that we tested, $p1$ was $1/2B$, while $p2$ and $p3$ were roughly $3/2B$ and $5/2B$, respectively, so the sampling rate should be chosen between $2B/3$ and $2B$. The actual value also depends on the pulse width of the input light, since $m$ must be an integer. However, a sampling rate between $2B/5$ and $2B/3$ will reduce the SNR enhancement. Excessive oversampling will not contribute to coding gain anymore.

Although oversampling has no effect in linear combination coding schemes like the Simplex code, they may also be affected by the bandwidth-limited feature of the PD, depending on the decoding process. However, the conclusions concerning coding gain may be different [10].

## 3. Experiments

Experiments were performed to verify the theoretical analysis, using the setup shown in Fig. 3. We employed a broadband amplified spontaneous emission (ASE) light source, operating at 1559 nm and with a 3 dB spectral width of 12 nm. The output power of continuous-wave (CW) light was set at 16 dBm. To modulate the CW light, it was used an acoustic optical modulator (AOM), which was controlled by an arbitrary function generator (AFG) that generated the code sequences or single pulses. The backscattered light was converted into an electrical signal by a 150 MHz highly sensitive photodetector (PD). Finally, the electrical signal was received in the oscilloscope and processed by a computer. The bandwidth of the AOM was 10 MHz, so the bit width or single pulse width could be no shorter than 100 ns. The maximum sampling rate of the oscilloscope was 20 gigasamples per second (GS/s), and the sampling rate could only be selected from a set of preset values. To ensure that the oversampling ratios were integer, in the following two experiments we chose 200 ns and 100 ns for the bit width and single pulse width, respectively. All the fibers under test were standard single-mode fibers (SMFs).

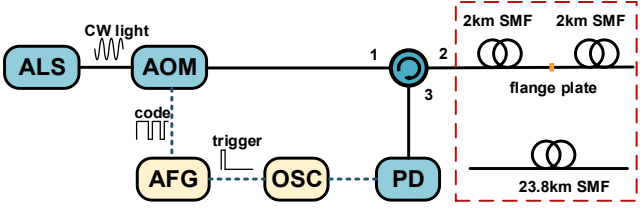

Fig. 3. Experimental setup for coded OTDR. ALS, amplified spontaneous emission light source; AOM, acoustic-optic modulator; AFG, arbitrary function generator; OSC, oscilloscope; PD, photodetector.

### A. Proving the triangular pulse phenomenon

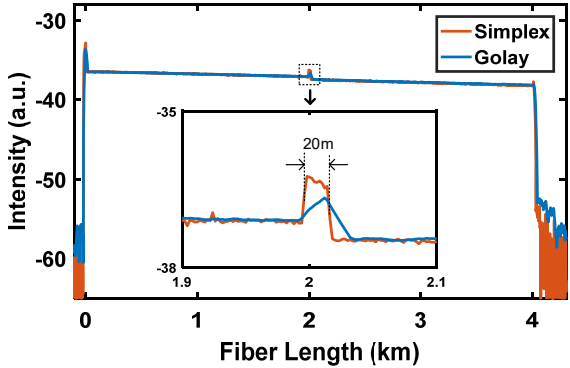

Fig. 4. Comparison between the intensities of the Golay-coded OTDR (blue) and Simplex-coded OTDR (red) response signals.

The goal of the first experiment was to prove that the equivalent probing pulse in correlation-coded OTDR with oversampling is a triangular pulse. Since the reflection peak in an OTDR curve should have the same shape as the probing pulse, we chose for our tests two 2 km SMFs connected by a flange plate carefully screwed to generate a small reflection in the junction. To reach a high enough SNR, the Simplex code was used to obtain a single pulse OTDR response The length of the Simplex and Golay codes was 255 and 2048, respectively. The pulse width was 200 ns, and the sampling rate was 50 MS/s in both cases, so that the oversampling ratio was 10. The results for the intensity as a function of the position in the fiber system are shown in Fig. 4. The two curves are almost identical, proving that the decoding process exactly recovered the OTDR response. Moreover, the shapes of the two reflection peaks are also in agreement with our analytical results.

### B. Coding gain verification

The second experiment was conducted to verify the coding gain analysis. Since the signal intensity is constantly changing along the fiber while the RMS of noise remains constant, the coding gain is obtained as follows:

$$\text{coding gain} = \frac{S_c/N_c}{S_p/N_p} = \frac{N_p}{N_c}, \quad (26)$$

where $S_c$ and $S_p$ are the signal intensities and $N_c$ and $N_p$ the noise intensities in the Golay-coded and single-pulse cases. As has been explained, $S_c$ and $S_p$ can be regarded as equivalent, so we only have to calculate the RMS noise in both cases. Since implementing the Golay code takes 4 measurements, the noise of the single-pulse measurement $N_p$ should be recorded 4 times and averaged for the comparison to be fair.

The fiber length in this case was 23.8 km, consuming 238 μs in a period of 500 μs. We chose 100 ns for the bit width, so the maximum code length could not be over 2048 bits, as the Golay code length must be a power of 2 [9]. The nonsignal section of the OTDR curve was used for calculating noise and further obtaining the coding gain through (26). The sampling rate ranged from 10 MS/s to 2.5 GS/s, corresponding to an oversampling ratio from 1 to 250. Golay sequences with code length ranging from 32 to 2048 were injected into the fiber, and each of the backscattered signals was acquired at the different sampling rates mentioned above.

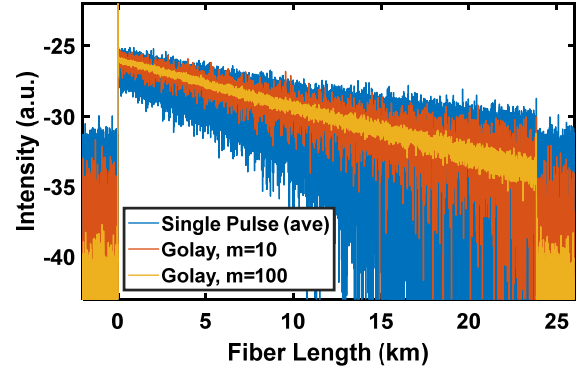

Fig. 5. Comparison between the intensities of the single-pulse OTDR (blue) and 2048 bit Golay-coded OTDR (red & yellow, for two different oversampling ratios) response signals. The single pulse result was averaged, while the coded results were not.

Fig. 5 shows a comparison between the intensities of the single-pulse and Golay-coded OTDR response signals. The blue curve gives the average single-pulse OTDR results calculated from 10000 measurements. We see that the signal can hardly be distinguished from the massive background noise. The red and yellow curves represent the results of 2048 bit Golay-coded OTDR with oversampling ratios of 10 and 100, respectively, without averaging. The SNR for the red curve is already better than for the blue curve, and that for the yellow curve is even better. Note that both coded results take 4 periods, or 2 ms, to complete one measurement, while the single pulse result takes 10000 periods, or 5 s, to acquire all data, and its SNR is still worse than in the coded cases. Implementing an averaging method in the coding scheme could serve to enhance the SNR even further.

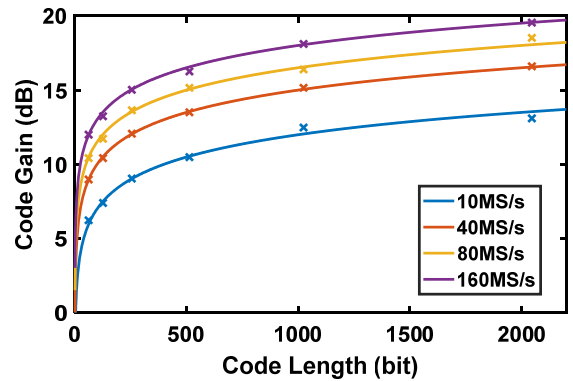

Fig. 6. Measured coding gain versus code length for different sampling rates. The curves represent theoretical values, while the crosses correspond to the experimental results.

Figure 6 shows the measured coding gain versus code length for different sampling rates. It is apparent that the measured results (represented by the crosses) agree perfectly with the theoretical curves. Note that for a sampling rate of 10 MS/s there is no oversampling, so Fig. 6 also proves that oversampling contributes to the SNR enhancement.

However, if we continue increasing the sampling rate, the experimental coding gain will diverge from the theoretical curve, as shown in Fig. 7. The values of the zero-crossings *p1* and *p2* for the PD used were measured to be 1.85 and 5.9 ns, respectively, as can be seen in Fig. 7(b). This means that the equivalent noise bandwidth was about 270 MHz and that the best sampling rate section in order to obtain extra SNR enhancement was from 180 to 540 MS/s. Obviously, the experimental results just agree with our conclusions. When the sampling rate was over 1 GS/s, the coding gain did not increase any more.

To make our analysis more convincing, we used another PD to repeat the experiment, and the results are shown in Fig. 8. This PD had a 300 MHz bandwidth, while its *p1*, *p2* and *p3* values were measured to be 0.9, 2.6, and 4.7 ns, respectively, as can be seen in Fig. 8(b). Thus, the equivalent noise bandwidth was about 550 MHz, and the preferred sampling rate section ranged from 370 to 1100 MS/s, while the "negative" sampling rate section (where the SNR is not enhanced, but reduced) went from 220 to 370 MS/s. Indeed, for this PD, when sampling rate is 250 MS/s, the coding gain will be lower than theoretical value, as shown in Fig. 8(a). The experimental results from these two PDs provide sufficient and convincing verification of our theoretical analysis.

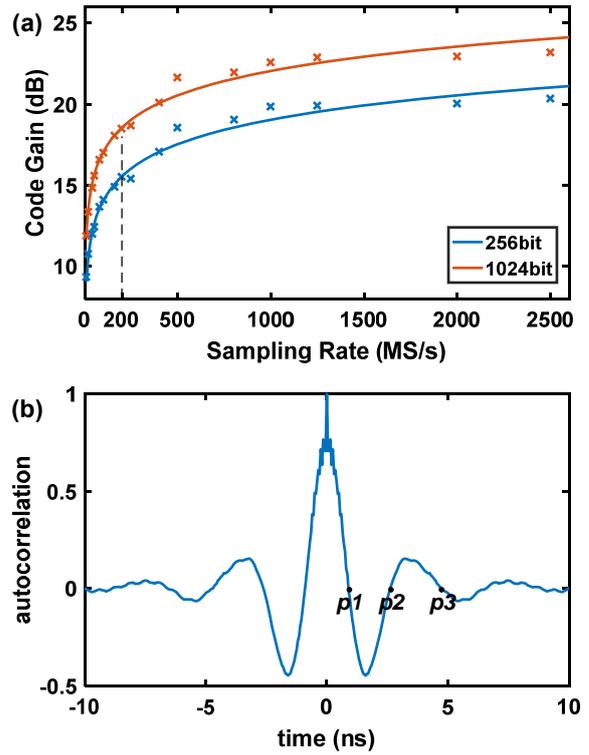

Fig. 8. (a) Measured coding gain versus sampling rate when the sampling rate becomes larger than the bandwidth of another 300 MHz PD. The curves represent theoretical values, while the crosses correspond to the experimental results. (b) Autocorrelation function of this PD noise.

## 4. Conclusion

We have made a thorough investigation of the sampling criteria for correlation-coded OTDR. Oversampling has been proven to further improve the coding gain without extending the measurement time, while it also turns the received OTDR curve into a triangular pulse response instead of the conventional, rectangular one. However, the coding gain is not always improved with the increase of the sampling rate due to the bandwidth-limited feature of the PD. One can take advantage of this feature to obtain extra SNR enhancement by choosing the sampling rate between *2B/3* and *2B*. To continue increasing the sampling rate would make no sense since the coding gain would reach a limit value. When the bit width of Golay codes is fixed, larger PD bandwidth can support larger oversampling ratio, resulting in better coding gain improvement. Therefore, large bandwidth PD is preferred in correlation-coded OTDR to fully utilize oversampling. Since oversampling does not require any change in the optical transmitting end, it can be conveniently applied to the existing systems.

Although the Golay code was taken as an example for our purposes, the sampling criteria proposed in this paper is also valid for other correlation coding schemes, such as the CCPONS code [8]. Moreover, this principle can also serve to significantly improve the performance of other fiber sensing systems based on direct detection OTDR, such as BOTDA and ROTDR.

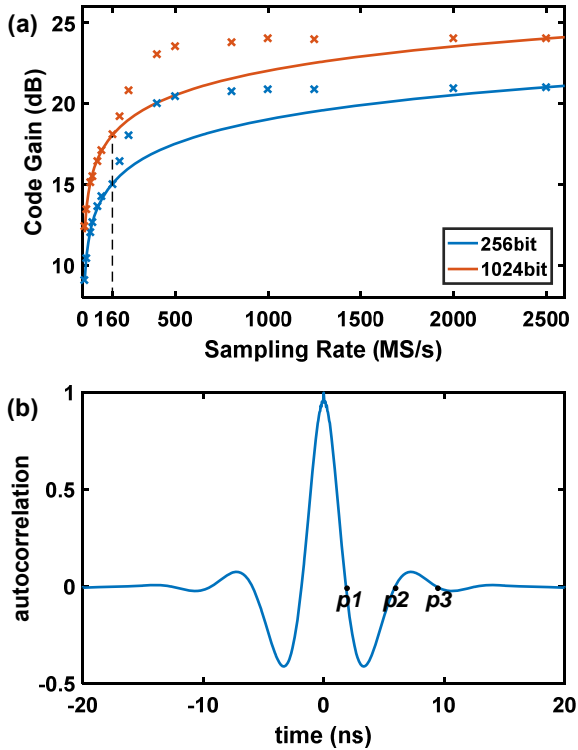

Fig. 7. (a) Measured coding gain versus sampling rate when the sampling rate becomes larger than the bandwidth of the 150 MHz PD. The curves represent theoretical values, while the crosses correspond to the experimental results. (b) Autocorrelation function of the PD noise.

**Funding.**
National Natural Science Foundation of China (61331010, 61290311); 863 High Technology Plan of China (2013AA013402);

**REFERENCES**

1. M. A. Soto, G. Bolognini, and F. Di Pasquale, "Analysis of optical pulse coding in spontaneous Brillouin-based distributed temperature sensors," Opt. Express **16**, 19097–19111 (2008).
2. M. D. Jones, "Using simplex codes to improve OTDR sensitivity," IEEE Photon. Technol. Lett. **5**, 822–824 (1993).
3. M. A. Soto, G. Bolognini, F. Di Pasquale, and L. Thévenaz, "Simplex-coded BOTDA fiber sensor with 1 m spatial resolution over a 50 km range," Opt. Lett. **35**, 259–261 (2010).
4. J. Park, G. Bolognini, D. Lee, P. Kim, P. Cho, F. Di Pasquale, and N. Park, "Raman-based distributed temperature sensor with simplex coding and link optimization," IEEE Photon. Technol. Lett. **18**, 1879–1881 (2006).
5. D. Lee, H. Yoon, P. Kim, J. Park, N. Y. Kim, and N. Park, "SNR enhancement of OTDR using biorthogonal codes and generalized inverses," IEEE Photon. Technol. Lett. **17**, 163–165 (2005).
6. M. Nazarathy, S. A. Newton, R. P. Giffard, D. S. Moberly, F. Sischka, W. R. Trutna, and S. Foster, "Real-time long range complementary correlation optical time domain reflectometer," J. Lightwave Technol. **7**, 24–38 (1989).
7. M. Nazarathy, S. A. Newton, and W. R. Trutna, "Complementary correlation OTDR with three codewords," Electron. Lett. **26**, 70–71 (1990).
8. P. K. Sahu, S. C. Gowre, and S. Mahapatra, "Optical time-domain reflectometer performance improvement using complementary correlated Prometheus orthonormal sequence," IET Optoelectronics **2**, 128–133 (2008).
9. M. Golay, "Complementary series," IRE Trans. Inf. Theory **7**, 82–87 (1961).
10. D. Lee, H. Yoon, N. Y. Kim, H. Lee, and N. Park, "Analysis and experimental demonstration of simplex coding technique for SNR enhancement of OTDR," in Proceedings of the Lightwave Technologies in Instrumentation and Measurement Conference (IEEE, 2004), pp. 118–122.